\documentclass[12pt,a4paper]{article}
\usepackage{amsmath,amssymb,mathtools,amsthm,mathrsfs,multirow,booktabs,morefloats,setspace,textcomp,epsfig}

\usepackage{caption,subfigure}

\usepackage[utf8]{inputenc}
\usepackage[section]{placeins}
\usepackage{lscape}
\usepackage{natbib}
\bibpunct{[}{]}{;}{n}{,}{,}

\newtheorem{theorem}{Theorem}

\newcounter{list-counter}
\newcounter{algo}[section]
\renewcommand{\thealgo}{\arabic{algo}}
\newenvironment{algo}
{
    \medskip
    \refstepcounter{algo}
    \noindent\textbf{Algorithm~\thealgo}
    \vspace{-7mm}
    \begin{list}{Step~\arabic{list-counter}.~}{\usecounter{list-counter}}{\medskip}
    \setlength{\itemsep}{-1mm}
}
{
    \end{list}
}

\title{\sc Bayesian Inference of a Dependent Competing Risk Data}
\date{}
\author{\sc Debashis Samanta{\thanks{D. Samanta is with the Department of Statistics and Informatics, Aliah University, II-A/27, Action Area II, Newtown, 
Kolkata, West Bengal, Pin 700156, 
W.B, India.}}, 
Debasis Kundu{\thanks{
{D. Kundu is with the Department of Mathematics and Statistics, IIT Kanpur, Pin 208016, India.  Corresponding
author, e-mail:kundu@iitk.ac.in}}}}
\date{}

\begin{document}
\maketitle
\parskip = 10pt
\doublespacing

\begin{center} 
{\sc Abstract}
\end{center}
Analysis of competing risks data plays an important role in the lifetime data analysis.  Recently Feizjavadian
and 
Hashemi (Computational Statistics and Data Analysis, vol. 82, 19-34, 2015) provided a classical inference of
a competing risks data set using four-parameter Marshall-Olkin bivariate Weibull 
distribution when the failure of an unit at a particular time point can happen due to more than one cause.    The aim of this paper is to provide the Bayesian analysis
of the same model based on a very flexible Gamma-Dirichlet prior on the scale parameters.  It is observed that
the Bayesian inference has certain advantages over the classical inference in this case.  We provide the Bayes estimates of the unknown parameters and the
associated highest posterior density credible intervals based on Gibbs sampling technique.  We further consider  the Bayesian inference of the model parameters assuming partially ordered Gamma-Dirichlet prior on the scale parameters when one cause is more severe than the other cause.  We have extended the results for different censoring schemes
also.


\section{\sc Introduction}
In lifetime data analysis an experimenter often wants to analyze data which have multiple failure modes.  In the 
statistical literature it is known as the competing risks problem.  There are mainly two different approaches to handle competing 
risks data.  One is known as the latent failure time model of Cox (1959) and the other is known as the cause specific hazard rate 
model of Prentice et al. (1978).  In case of exponential and Weibull lifetime distributions it has been
shown by Kundu (2004) that both the models lead to the same likelihood function, although their
interpretations are different.
An extensive list of literature exits in this area, see for
example Kalbfleish and Prentice (1980), Lawless (1982) or Crowder (2001) and the references cited therein.  Most of the existing studies are based on the assumptions that  the causes of failures are
independent, although it may not be 
true in practice.  It may be mentioned that there are some identifiability issues in this respect, see for
example Tsiatis (1975).

The bivariate or multivariate lifetime distributions play an important role in analyzing dependent competing risk model.  Bayesian inference of a dependent competing risk model assuming absolute continuous bivariate exponential distribution is studied by Wang and Ghosh (2003).  When there is a positive probability of simultaneous occurrence of two causes of failure then Marshall-Olkin bivariate exponential (MOBE) distribution, introduced by
Marshall and Olkin (1967), can be used to analyze the data.  If the data indicate that the marginals have unimodal probability density functions (PDFs) then MOBE distribution will not fit the data.  Due to this limitations, Marshall-Olkin bivariate Weibull (MOBW) distribution was introduced by Lu (1989).  Later this distribution has been studied by several authors including Jose, Ristic and Joseph (2011), Dey and Kundu (2009) and
Kundu and Gupta (2013).  The analysis of dependent competing risk model using MOBW distribution is considered by
Feizjavdian and Hashemi (2015).  Bayesian inference of a series system with dependent causes of failure using MOBW distribution is provided by Xu and Zhou (2017).  Different methods of estimating parameters of dependent competing risks using MOBW model has been studied by Shen and Xu (2018).

Order restriction among model parameters in a reliability model has been considered by several authors.  Order restricted inference of step-stress model has been considered by
Balakrishnan, Beutner and Kateri (2009) and Samanta and Kundu (2018) to incorporate the fact that
the increased stress level will reduce the expected lifetime of the experimental units. Recently Mondal and
Kundu (2020) considered order restricted inference  for
two exponential populations.  They have mentioned that this order restricted inference can be used in an accelerated life test if one sample is put under higher stress keeping the other one in
normal stress.  In competing risk model when it is known apriori that one cause of failure is higher risk than the other then we may incorporate this information by considering an order restriction on the model parameters.

The motivation of this paper came from a recent paper by Feizjavdian and Hashemi (2015).  They have
analyzed a data set obtained from the Diabetic Retinopathy Study (DRS) conducted by the National Eye Institute to
estimate the effect of laser treatment in delaying the onset of blindness in patients with diabetic retinopathy.  At
the beginning of the experiment, for each patient, one eye was selected for laser treatment and the other eye was
not given the laser treatment.  For each patient the minimum time to blindness ($T$) and the indicator specifying
whether the treated eye ($\delta$ = 1) or the untreated eye ($\delta$ = 2) has first failed has been recorded.
If both the eyes have failed simultaneously then $\delta$ = 0 has been recorded.  The data set is presented
in Table \ref{tab:data1}.  The main objective of this experiment is to study whether the laser treatment has any effect
on delaying the onset of blindness in patients with diabetic retinopathy.
Clearly, the time to blindness of the two eyes cannot be independent and there are some ties
in the data set.  Due to this reason Feizjavdian and Hashemi (2015) considered a dependent competing risks
model and they have proposed to use the Marshall-Olkin bivariate
Weibull distribution for this purpose.  They provided the maximum likelihood estimators (MLEs) of the unknown parameters and 
obtained the associated asymptotic confidence intervals.  The maximum likelihood estimators cannot be obtained in explicit forms, hence
they have obtained the approximate maximum likelihood estimators which can be obtained in explicit forms.
It is observed that the proposed model works quite well for fitting purposes.  They have observed that even for
highly censored data, the MLEs perform quite well.

The main aim of this paper is to provide the Bayesian analysis of the same data set  under a very flexible
Gamma-Dirichlet (GD) prior 
on the scale parameters and for a very general log-concave prior on the shape parameter.  It is observed that the
Bayesian inference has some natural advantages in this case.
The Gamma-Dirichlet prior was originally 
introduced by Pena and Gupta (1990) for Marshall-Olkin bivariate exponential distribution (MOBE).  The GD prior is a very flexible 
prior, and its joint PDF can take variety of 
shapes depending on the hyper parameters.  It can be both positively and negatively correlated.  In case of 
MOBW distribution the GD distribution is a conjugate prior of the scale parameters for a fixed 
shape parameter.  Hence the posterior distribution of the scale parameters for the fixed shape parameter can be obtained in a very 
convenient form.  We have used a very general log-concave prior on the shape parameter, and they are assumed to be independent.  
The Bayes estimators cannot be obtained in 
closed form.  We have used Gibbs sampling technique to compute the Bayes estimates and the associated highest posterior density (HPD)
credible intervals.  

We further consider the Bayesian inference of the model parameters with partially order restriction on scale parameters.  This order restriction comes naturally when it is apriori known that one cause of failure is more severe than the other.  In this case we consider partially order restricted Gamma-Dirichlet prior for scale parameters and we use importance sampling technique for Bayes estimates and the credible intervals. We re-analyze the same data set assuming the order restriction on scale parameters.  One major advantage of the Bayesian inference is that different forms of data for example; Type-I, Type-II, 
hybrid censored data can be handled quite conveniently also, unlike the classical inference.
Finally the Bayesian testing of hypothesis has been considered to test the hypothesis that there is no significant difference between two causes of failure.  We propose to use Bayes factor to test the hypothesis and interestingly
in this case it can be obtained in explicit form.  We have reanalyzed the data set and it is observed that the
laser treatment does not have any effect in delaying the onset of blindness.

The rest of the paper is organized as follows.  In Section 2 we describe the Marshall-Olkin bivariate Weibull distribution and provide the likelihood function based on competing risk data.  Prior assumptions and posterior analysis is provided in Section 2.1.  The order restricted Bayesian inference is given in Section 3.  The inference under different censoring scheme is provided in Section 4.  In Section 5, we discuss the Bayesian testing of hypothesis problem.
The analysis of the data set has been provided in Section 6 and 
an extensive simulation results have been discussed in Section 7.  Finally we have concluded the article in Section
8.

\section{\sc Marshal-Olkin Bivariate Weibull Competing Risk Model}
Suppose a life testing experiment starts with $n$ number of identical units at time zero and the failure times are recorded.  We assume that the units are failed due to several causes of failure.  Here we restrict ourselves to
two causes of failure, although the results can be easily generalized for more than two causes also.
Let $X_1$ be a random variable associated with the lifetime of an unit under the first cause and $X_2$ be a random 
variable associated with the second cause.  An unit is failed if the minimum of the two occurs.  Therefore 
$T=min\{X_1,X_2\}$ is the random variable associated with the lifetime of an experimental unit which is exposed
to both the risk factors.  Along with the failure time, the cause of failure is also recorded.  In reality the two causes are related and hence we assume MOBW 
distribution for two causes of failure.  In this model it is assumed that the failure can occur due to both the causes simultaneously.  The MOBW distribution is defined as follows: 
Suppose for $i=0,1,2$,  $U_i$ follows independent Weibull distribution with shape parameter $\alpha$ and 
scale parameter $\lambda_i$ .  We will denote it by $U_i\sim WE(\alpha,\lambda_i)$.  The PDF and the survival function of $U_i$ for $u_i > 0$ are, respectively,  
\begin{eqnarray}
 f_{WE}(u_i;\,\alpha,\,\lambda_i) = \alpha \lambda_i u_i^{\alpha-1} e^{-\lambda_i u_i^{\alpha}} \ \ \ \ \ \ \text{and} \ \ \ \ \ \ S_{WE}(u_i;\,\alpha,\,\lambda_i) =  e^{-\lambda_i u_i^{\alpha}}.  \nonumber  
\end{eqnarray}
Now let $X_1 = min\{U_0,\,U_1\}$ and $X_2 = min\{U_0,\,U_2\}$, then the bivariate random variable $(X_1,X_2)$ 
is said to be follow the MOBW distribution with shape parameter $\alpha$ and scale parameters $\lambda_0$, $\lambda_1$ 
and $\lambda_2$ and it is denoted by $(X_1,X_2) \sim MOBW(\alpha,\lambda_0,\lambda_1,\lambda_2)$.  The survival 
function of MOBW random variable $(X_1,X_2)$ is 
\begin{align} \label{survai1}
        S_{X_1,X_2}(x_1,x_2) = \left\{ \begin{array}{lll}
	                        S_{WE}(x_1;\,\alpha,\,\lambda_1) S_{WE}(x_2;\,\alpha,\,\lambda_0+\lambda_2) \quad & \text{if} & \ \ \ x_1 < x_2 \\ 
                                S_{WE}(x_1;\,\alpha,\,\lambda_0+\lambda_1) S_{WE}(x_2;\,\alpha,\,\lambda_2) \quad & \text{if} & \ \ \ x_1 > x_2 \\ 
                                S_{WE}(x;\,\alpha,\,\lambda_0+\lambda_1+\lambda_2) \quad & \text{if} &\ \ \ x_1 = x_2=x.
                    \end{array}
                \right.
\end{align}
The joint PDF of $(X_1,X_2)$ is 
\begin{align} \label{pdf1}
        f_{X_1,X_2}(x_1,x_2) = \left\{ \begin{array}{lll}
	                        f_1(x_1,x_2) \quad & \text{if} & \ \ \ x_1 < x_2 \\ 
                                f_2(x_1,x_2) \quad & \text{if} & \ \ \ x_1 > x_2 \\ 
                                f_0(x,x) \quad & \text{if} &\ \ \ x_1 = x_2=x ,
                    \end{array}
                \right.
\end{align}
where
\begin{eqnarray}
 f_1(x_1,x_2) & = & f_{WE}(x_1;\,\alpha,\,\lambda_1) f_{WE}(x_2;\,\alpha,\,\lambda_0+\lambda_2),   \nonumber  \\
 f_2(x_1,x_2) & = & f_{WE}(x_1;\,\alpha,\,\lambda_0+\lambda_1) f_{WE}(x_2;\,\alpha,\,\lambda_2),  \nonumber  \\
 f_0(x,x) & = & \frac{\lambda_0}{\lambda_0+\lambda_1+\lambda_2}f_{WE}(x;\,\alpha,\,\lambda_0+\lambda_1+\lambda_2).   \nonumber
\end{eqnarray}

Let us define 
\begin{align} 
        \Delta = \left\{ \begin{array}{lll}
	                        0 \quad & \text{if} & \ \ \ \text{the failure occur due to both the causes simultaneously} \\ \nonumber
                                1 \quad & \text{if} & \ \ \ \text{the failure occur due to first cause} \\ \nonumber
                                2 \quad & \text{if} &\ \ \ \text{the failure occur due to second cause.}
                    \end{array}
                \right.
\end{align}
Suppose we observe the failure time of the experimental units along with the cause of failure.  Therefore, $T = min\{X_1, X_2\}$ and $\Delta$ be the random variables corresponding to the failure time and the cause of failure of an experimental unit respectively.  Thus the available data set on a competing risk model is of the form: 
$\{(t_{1:n},\delta_1),\ldots,(t_{n:n},\delta_n)\},$ where $(t_{i:n},\delta_i)$ denotes the $i$-th ordered observed value of $(T,\Delta)$.
The likelihood function based on the data set can be obtained using the below equation
\begin{eqnarray}\label{likeli1}
 L(\alpha,\lambda_0,\lambda_1,\lambda_2 \vert Data) & \propto & \prod_{i=1}^{n}[f_{X_1,X_2}(t_{i:n},t_{i:n})]^{\delta_{i0}} \bigg[-\frac{\partial}{\partial x_1} S_{X_1,X_2}(x_1,x_2)\mid_{(t_{i:n},t_{i:n})}\bigg]^{\delta_{i1}}  \nonumber \\
 {} & {} & \bigg[-\frac{\partial}{\partial x_2} S_{X_1,X_2}(x_1,x_2)\mid_{(t_{i:n},t_{i:n})}\bigg]^{\delta_{i2}},
\end{eqnarray}
where $\delta_{i0}$, $\delta_{i1}$, $\delta_{i2}$ are the indicators for failure of $i$-th observation due to both the causes simultaneously, first cause and 
second cause respectively.  From the survival function in (\ref{survai1}) of MOBW distribution we have
\begin{align}
\begin{array}{lll} \label{survai2}
 -\frac{\partial}{\partial x_1} S_{X_1,X_2}(x_1,x_2)\mid_{(t_{i:n},t_{i:n})} & = & f_{WE}(t_{i:n};\alpha,\lambda_1) S_{WE}(t_{i:n}; \alpha, \lambda_0+\lambda_2) , \\ 
 -\frac{\partial}{\partial x_2} S_{X_1,X_2}(x_1,x_2)\mid_{(t_{i:n},t_{i:n})} & = & S_{WE}(t_{i:n}; \alpha, \lambda_0+\lambda_1) f_{WE}(t_{i:n};\alpha,\lambda_2) . 
\end{array}
\end{align}
Therefore using (\ref{likeli1}) and (\ref{survai2}) the likelihood of the data is
\begin{align}
 \begin{array}{lll} \label{likeli2}
  L(\alpha,\lambda_0,\lambda_1,\lambda_2 \vert Data) & \propto & \alpha^n \lambda_0^{n_0} \lambda_1^{n_1} \lambda_2^{n_2} \bigg( \prod_{i=1}^{n} t_{i:n}^{\alpha-1} \bigg ) 
  e^{-(\lambda_0+\lambda_1+\lambda_2)\sum_{i=1}^{n} t_{i:n}^{\alpha}},
 \end{array}
\end{align}
where $n_0 = \sum_{i=1}^{n} \delta_{i0}$, $n_1 = \sum_{i=1}^{n} \delta_{i1}$ and $n_2 = \sum_{i=1}^{n} \delta_{i2}$ ($n_i>0$ for $i = 0,1,2$ 
and $n =\sum_{i=0}^{2} n_i$) are the number of failures due to both the causes, the first cause and the second cause, respectively.

\subsection{\sc Prior Assumption and Posterior Analysis}
In this section we will provide the Bayesian inference of the model parameters under squared error loss function.  
Since we have considered 
a dependent competing risk model, in the Bayesian analysis we assume a dependent prior distribution of $(\lambda_0, \lambda_1, 
\lambda_2)$.  Using the concept of Pena and Gupta (1990) we have assumed the multivariate Gamma-Dirichlet prior for $(\lambda_0, \lambda_1, 
\lambda_2)$.  Therefore the joint prior distribution of $(\lambda_0, \lambda_1, \lambda_2)$ with hyper parameters $a>0$, $b>0$, $a_0>0$, 
$a_1>0$ and $a_2>0$ is given by
\begin{eqnarray} \label{prior1}
\pi_0 (\lambda_0, \lambda_1, \lambda_2 \vert a, b, a_0, a_1, a_2) &  = &  \frac{\Gamma{(\overline{a})}}{\Gamma{(a)}}(b\lambda)^{a-\overline{a}} 
\prod_{i=0}^{2} \frac{b^{a_i}}{\Gamma{(a_i)}} \lambda_i^{a_i-1} e^{-b\lambda_i}, 
\end{eqnarray}
where $\overline{a}=a_0+a_1+a_2$ and $\lambda = \lambda_0+\lambda_1+\lambda_2$.  This distribution will be denoted by 
$GD(a,b,a_0,a_1,a_2)$.  In general this is a dependent prior but if $a=\overline{a}$ then $\lambda_i$'s are 
independent gamma priors with parameter $b$ and $a_i$ ($i=0,1,2$).  The prior 
distribution of $\alpha$ is Gamma with hyper parameters $c_1>0$ and $c_2>0$ (denoted by $GA(c_1,c_2)$) and is independent with the joint prior distribution 
of $(\lambda_0,\lambda_1,\lambda_2)$.  Thus the joint prior of $(\alpha,\lambda_0,\lambda_1,\lambda_2)$ is given by 
\begin{eqnarray} \label{prior2}
\pi_1 (\alpha,\lambda_0, \lambda_1, \lambda_2 \vert a, b, a_0, a_1, a_2,c_1,c_2) &  = &  \frac{c_1^{c_2}}{\Gamma(c_2)} e^{-c_1\alpha} \alpha^{c_2-1} \times \frac{\Gamma(\overline{a})}{\Gamma(a)}(b\lambda)^{a-\overline{a}} 
\prod_{i=0}^{2} \frac{b^{a_i}}{\Gamma(a_i)} \lambda_i^{a_i-1} e^{-b\lambda_i}.  \nonumber \\
\end{eqnarray}
Therefore the joint posterior distribution of $(\alpha,\lambda_0, \lambda_1, \lambda_2)$ is given by
\begin{align}
 \begin{array}{lll} \label{posteri2}
  \widetilde{\pi}(\alpha, \lambda_0,\lambda_1,\lambda_2 \vert Data) & \propto & \widetilde{\pi_1}(\alpha) \widetilde{\pi_2}(\lambda_0,\lambda_1,\lambda_2 \vert \alpha),
 \end{array}
\end{align}
where, 
\begin{align}
 \begin{array}{lll}
  \widetilde{\pi_1}(\alpha) & = & e^{-c_1 \alpha} \alpha^{n+c_2-1} \big[b+\sum_{i=1}^{n} t_{i:n}^{\alpha} \big]^{-(a+n)} \prod_{i=1}^{n} t_{i:n}^{\alpha-1}, \nonumber \\
  \widetilde{\pi_2}(\lambda_0,\lambda_1,\lambda_2 \vert \alpha) & = & \frac{\Gamma(\overline{a}+n)}{\Gamma{(a+n)}}
  \big[\{b+\sum_{i=1}^{n} t_{i:n}^{\alpha} \} \lambda \big]^{[(a+n)-(\overline{a}+n)]} \\
  {} & {} & \times \prod_{j=0}^{2} \frac{\big[b+\sum_{i=1}^{n} t_{i:n}^{\alpha}\big]^{a_j+n_j}}{\Gamma(a_j+n_j)} 
  \lambda_j^{a_j+n_j-1} e^{-\lambda_j\big[b+\sum_{i=1}^{n} t_{i:n}^{\alpha}\big]}.
 \end{array}
\end{align}

In this case the explicit form of the Bayes estimates cannot be obtained and hence we propose to use 
Gibbs sampling technique to obtain the Bayes estimates and associated credible intervals.  The form of the $\widetilde{\pi_1}(\alpha)$ is not any standard distributional form but in Theorem (\ref{theo:logcncav1}) we will show 
that $\widetilde{\pi_1}(\alpha)$ is a log-concave density function.  On the other hand, for a given $\alpha$ the joint posterior 
distribution of $(\lambda_0,\lambda_1,\lambda_2)$, i.e., $\widetilde{\pi_2}(\lambda_0,\lambda_1,\lambda_2 \vert \alpha)$ is 
$GD(a+n,b+\sum_{i=1}^{n} t_{i:n}^{\alpha}, a_0+n_0, a_1+n_1, a_2+n_2)$.
\begin{theorem}\label{theo:logcncav1}
$\widetilde{\pi_1}(\alpha)$ is a log-concave density function.
\begin{proof}
 See in the Appendix.  
\end{proof}
\end{theorem}
The method 
proposed by Devroye (1984) for generation of random sample from a log-concave density function can be 
used to generate sample from $\widetilde{\pi_1}(\alpha)$.  Generation of sample from Gamma-Dirichlet distribution is 
quite straight forward which is given explicitly in Kundu and Pradhan (2011).  Thus we propose to execute the 
following algorithm to obtain the Bayes estimates and the associated credible intervals of the 
unknown parameters.

\begin{algo}
 \item Generate $\alpha$ from $\widetilde{\pi_1}(\alpha)$ using the method proposed by Devroye (1984)
or the ratio-of-uniform method introduced by Kinderman and Monahan (1977).
   \item For a given $\alpha$ generate $(\lambda_0,\,\lambda_1,\,\lambda_2)$ from $GD(a+n,b+\sum_{i=1}^{m} t_{i:n}^{\alpha}, a_0+n_0, a_1+n_1, a_2+n_2)$.
   \item Repeat Step 1 and Step 2, $M$ times to obtain $(\alpha^1,\lambda_0^1,\lambda_1^1,\lambda_2^1, \ldots, \alpha^M,\lambda_0^M,\lambda_1^M,\lambda_2^M)$.  
   \item Bayes estimate of $\alpha$, $\lambda_0$, $\lambda_1$ and $\lambda_2$ with respect
   to squared error loss function are respectively given by
\begin{eqnarray}
\displaystyle  \widehat{\alpha}_{(B)} = \frac{1}{M}\sum_{k=1}^{M} \alpha^k, & \displaystyle \widehat{\lambda}_{0(B)} = \displaystyle \frac{1}{M}\sum_{k=1}^{M} \lambda_{0}^k,   \quad 
 \widehat{\lambda}_{1(B)} = \frac{1}{M}\sum_{k=1}^{M} \lambda_{1}^k, & \widehat{\lambda}_{2(B)} = \frac{1}{M}\sum_{k=1}^{M} \lambda_{2}^k. \nonumber
\end{eqnarray}
\item The corresponding posterior variance can be obtained respectively as
\begin{eqnarray}
 V_{post}(\alpha) = \frac{1}{M}\sum_{k=1}^{M} (\alpha^k-\widehat{\alpha}_{(B)})^2, & \quad \displaystyle V_{post}(\lambda_0) = \frac{1}{M}\sum_{k=1}^{M} (\lambda_0^k-\widehat{\lambda}_{0(B)})^2, \nonumber \\ 
 V_{post}(\lambda_1) = \frac{1}{M}\sum_{k=1}^{M} (\lambda_1^k-\widehat{\lambda}_{1(B)})^2, & \quad \displaystyle V_{post}(\lambda_2) = \frac{1}{M}\sum_{k=1}^{M} (\lambda_{2}^k-\widehat{\lambda}_{2(B)})^2. \nonumber 
\end{eqnarray}
\item To obtain credible interval of $\alpha$, we order $\alpha^{1},\ldots,\alpha^{M}$ as $\alpha^{(1)}<\ldots<\alpha^{(M)}$.
Then $100(1-\gamma)\%$ symmetric credible interval of $\alpha$ is given by $(\alpha^{([\frac{\gamma}{2}M])},\alpha^{([(1-\frac{\gamma}{2})M])}).$
\item To construct $100(1-\gamma)\%$ highest posterior density (HPD) credible interval of $\alpha$, consider the set of credible intervals 
$(\alpha^{(j)},\alpha^{([j+(1-{\gamma})M])})$, $j=1,\ldots,[\gamma M]$.  Therefore  $100(1-\gamma)\%$ HPD credible interval of $\alpha$
is $(\alpha^{(j^*)},\alpha^{([j^*+(1-{\gamma})M])})$, where $j^*$ is such that 
\begin{align}
 \alpha^{([j^*+(1-{\gamma})M])} - \alpha^{(j^*)} < \alpha^{([j+(1-{\gamma})M])} - \alpha^{(j)} \quad \text{for all} \quad j= 1 \ldots [\gamma M]. \nonumber
\end{align}
Similar to Step 6 and Step 7 we can obtain the symmetric and HPD credible intervals for other parameters.
 \end{algo}

\section{\sc Order Restricted Inference}
In this section we provide the order restricted Bayesian inference of the model parameters.  Between two causes, let cause - 1 be more severe than cause - 2.  Therefore, there is a ordering between the parameters related to two causes.  In this model assumption, the ordering is $\lambda_1 < \lambda_2$.  We want to incorporate this information in our inference.  In order restricted inference, we consider the following joint prior distribution of $(\lambda_0,\lambda_1,\lambda_2)$ assuming $\lambda_1 < \lambda_2$.  Let
 \begin{eqnarray} \label{prior3}
\pi_0 (\lambda_0, \lambda_1, \lambda_2 \vert a, b, a_0, a_1, a_2) &  = &  \frac{\Gamma(\overline{a})}{\Gamma(a)}(b\lambda)^{a-\overline{a}} 
\prod_{i=0}^{2} \frac{b^{a_i}}{\Gamma(a_i)} \lambda_0^{a_0-1} e^{-b\lambda} (\lambda_1^{a_1-1}\lambda_2^{a_2-1} + \lambda_2^{a_1-1}\lambda_1^{a_2-1}). 
\end{eqnarray}
Note that the above prior distribution is the joint PDF of partially ordered random variables $(\lambda_0,\lambda_{(1)},\lambda_{(2)})$, where $(\lambda_0,\lambda_{(1)},\lambda_{(2)}) = (\lambda_0,\lambda_1,\lambda_2)$ if  $\lambda_1 < \lambda_2$ and $(\lambda_0,\lambda_{(1)},\lambda_{(2)}) = (\lambda_0,\lambda_2,\lambda_1)$ if  $\lambda_2 < \lambda_1$ and $(\lambda_0, \lambda_1, \lambda_2) \sim GD(a,b,a_0,a_1,a_2)$.   We denote the prior in (\ref{prior3}) as $POGD(a, \, b, \, a_0, \,a_1, \, a_2)$.  Here also we assume that the prior 
distribution of $\alpha$ is Gamma with hyper parameters $c_1>0$ and $c_2>0$ and is independent with the joint prior distribution 
of $(\lambda_0,\lambda_1,\lambda_2)$.  The explicit form of the Bayes estimates under squared error loss function cannot be obtained.  Hence we propose to use importance sampling technique to obtain the Bayes estimates and the associated credible intervals.  The joint posterior distribution can be written as
\begin{align}
 \begin{array}{lll} \label{posteri3}
  \widetilde{\pi}(\alpha, \lambda_0,\lambda_1,\lambda_2 \vert Data) & \propto & \widetilde{\pi_1}(\alpha) \widetilde{\pi_2}(\lambda_0,\lambda_1,\lambda_2 \vert \alpha) h(\alpha, \lambda_0,\lambda_1,\lambda_2),
 \end{array}
\end{align}
where
\begin{align}
 \begin{array}{lll} 
  \widetilde{\pi_1}(\alpha) & \propto & e^{-c_1 \alpha} \alpha^{n+c_2-1} \big[b+\sum_{i=1}^{n} t_{i:n}^{\alpha} \big]^{-(a+n)} \prod_{i=1}^{n} t_{i:n}^{\alpha-1}, \nonumber \\
  \widetilde{\pi_2}(\lambda_0,\lambda_1,\lambda_2 \vert \alpha) & = & \frac{\Gamma(\overline{a}+2n)}{\Gamma{(a+n)}\Gamma{(a_0+2n_0)}\Gamma{(a_1+n_1+n_2)}\Gamma{(a_2+n_1 + n_2)}}
  \big[\{b+\sum_{i=1}^{n} t_{i:n}^{\alpha} \} \lambda \big]^{[(a+n)-(\overline{a}+2n)]} \\
  {} & {} & \times \big[b+\sum_{i=1}^{n} t_{i:n}^{\alpha}\big]^{\overline{a}+2n} 
  \lambda_0^{a_0+2n_0-1} e^{-\lambda\big[b+\sum_{i=1}^{n} t_{i:n}^{\alpha}\big]} \\
	{} & {} & \times \big[ \lambda_1^{a_1+n_1 + n_2-1}\lambda_2^{a_2+n_1 + n_2-1} + \lambda_1^{a_2+n_1 + n_2-1}\lambda_2^{a_1+n_1 + n_2-1}\big], \\
h(\alpha, \lambda_0,\lambda_1,\lambda_2) & = & 	\frac{\lambda^n}{\lambda_0^{n_0} \lambda_1^{n_2} \lambda_2^{n_1}} .
 \end{array}
\end{align}

\noindent As before $\widetilde{\pi_1}(\alpha)$ is a log-concave density function and hence we can generate $\alpha$ from $\widetilde{\pi_1}(\alpha)$ easily.  Also note that $\widetilde{\pi_2}(\lambda_0,\lambda_1,\lambda_2 \vert \alpha)$ is POGD($a+n, b+\sum_{i=1}^{n} t_{i:n}^{\alpha}, a_0+2n_0, a_1+n_1+n_2, a_2+n_1+n_2$) and generation from this distribution is quite straight forward.  For given $\alpha$, first generate $(\lambda_0^*,\lambda_1^*,\lambda_2^*)$ from  GD($a+n, b+\sum_{i=1}^{n} t_{i:n}^{\alpha}, a_0+2n_0, a_1+n_1+n_2, a_2+n_1+n_2$) and then take $(\lambda_0,\lambda_1,\lambda_2) = (\lambda_0^*,\lambda_1^*,\lambda_2^*)$ if $\lambda_1^*<\lambda_2^*$ otherwise if $\lambda_2^*<\lambda_1^*$ then take $(\lambda_0,\lambda_1,\lambda_2) = (\lambda_0^*,\lambda_2^*,\lambda_1^*)$.  Now we propose the following algorithm for Bayes estimates and the associated credible intervals.

\noindent {\bf Algorithm 2:}

\noindent Step 1: Generate $\alpha_1$ from $\widetilde{\pi_1}(\alpha)$ using the method proposed by
Devroye (1984) or the ratio-of-uniform method introduced by Kinderman and Monahan (1977).

\noindent   Step 2. For a given $\alpha_1$ generate $(\lambda_{01},\,\lambda_{11},\,\lambda_{21})$ from $POGD(a+n, b+\sum_{i=1}^{m} t_{i:n}^{\alpha_1}, a_0+2n_0, a_1+n_1+n_2, a_2+n_1+n_2)$.
	
\noindent Step 3: Repeat Step 1-Step 2, $M$ times to get $(\alpha_1, \lambda_{01}, \lambda_{11}, \lambda_{21}),$ $\ldots,$ $(\alpha_M, \lambda_{0M}, \lambda_{1M}, \lambda_{2M})$.    

\noindent Step 4: Compute $g_i = g(\alpha_i, \lambda_{0i}, \lambda_{1i}, \lambda_{2i}) ; i = 1,\ldots,M$.

\noindent Step 5: Calculate the weights $w_{i}=\frac{h(\alpha_i, \lambda_{0i}, \lambda_{1i}, \lambda_{2i})}{\sum_{i=1}^{M}h(\alpha_i, \lambda_{0i}, \lambda_{1i}, \lambda_{2i})}$.

\noindent Step 6: Compute the BE of $g(\alpha, \lambda_0, \lambda_1, \lambda_2)$ under the squared error loss function as  \linebreak
$\widehat{g}_B(\alpha, \lambda_0, \lambda_1, \lambda_2) = \sum_{j=1}^{M}w_{j}g_j$.

\noindent Step 7: To construct a $100(1-\gamma)\%$ $(0<\gamma<1)$ CRI of $g(\alpha, \lambda_0, \lambda_1, \lambda_2),$ first order
$g_j's$ for j=1,2,\ldots, M, say $g_{(1)}<g_{(2)}<\ldots<g_{(M)}$ and arrange $w_{j}$ accordingly to get $w_{(1)}, w_{(2)},\ldots,w_{(M)}.$
Note that $w_{(1)}, w_{(2)},\ldots,w_{(M)}$ may not be ordered. 

\noindent Step 8: A $100(1-\gamma)\%$ CRI can be obtain as $(g_{j_1},g_{j_2})$ where $j_1$ and $j_{2}$ satisfy 
\begin{align}\label{eq:hpdcri}
j_1,\,j_2\in\left\{1,\,2,\,\ldots,\,M\right\}, \quad j_1<j_2, \quad\sum_{i=j_1}^{j_2}w_{(i)}\leq 1-\gamma<\sum_{i=j_1}^{j_2+1}w_{(i)}.
\end{align}
The $100(1-\gamma)\%$ HPD CRI of $g(\alpha, \lambda_0, \lambda_1, \lambda_2)$ becomes
$\left(g_{(j_1^*)},\,g_{(j_2^*)}\right)$, where $1\leq j_1^*<j_2^*\leq M$ satisfy
\begin{align*}
\sum_{i=j_1^*}^{j_2^*}w_{(i)}\leq 1-\gamma<\sum_{i=j_1^*}^{j_2^*+1}w_{(i)}, \quad\text{and}\quad
g_{(j_2^*)}-g_{(j_1^*)}\leq g_{(j_2)}-g_{(j_1)},
\end{align*}
for all $j_1$ and $j_2$ satisfying \eqref{eq:hpdcri}.

\section{\sc Inference under Different Censoring Schemes}
There are several censoring schemes available in the literature.  One major advantage of the Bayesian
inference is that we can easily extend the inference to different censoring schemes.  In this section we discuss the inference of dependent competing risk model under different censoring schemes.  Before proceeding, we define the following notations.  
$\tau^* = $ termination time of the experiment; $n^* = $ total number of failure before $\tau^*.$

\subsection{\sc Type-I Censoring}
In Type-I censoring scheme we stop the experiment at a prefix time, say $\tau^*$ and the number of observations failed before $\tau^*$ is $n^*$. In this case observed data is of the form 
 $\{(t_{1:n},\delta_1),\ldots,(t_{n^*:n},\delta_{n^*}) \}$.   
In this case the likelihood of the data is given by
\begin{eqnarray}\label{likeli3}
 L(\alpha,\lambda_0,\lambda_1,\lambda_2 \vert Data) & \propto & \prod_{i=1}^{n^*}[f_{X_1,X_2}(t_{i:n},t_{i:n})]^{\delta_{i0}} \bigg[-\frac{\partial}{\partial x_1} S_{X_1,X_2}(x_1,x_2)\mid_{(t_{i:n},t_{i:n})}\bigg]^{\delta_{i1}}  \nonumber \\
 {} & {} & \bigg[-\frac{\partial}{\partial x_2} S_{X_1,X_2}(x_1,x_2)\mid_{(t_{i:n},t_{i:n})}\bigg]^{\delta_{i2}} \bigg[S_{X_1,X_2}(x_1,x_2)\mid_{(\tau^*,\tau^*)}\bigg]^{n-n^*} \nonumber \\
{} & = & \alpha^{n^*} \lambda_0^{n_0} \lambda_1^{n_1} \lambda_2^{n_2} \bigg( \prod_{i=1}^{n^*} t_{i:n}^{\alpha-1} \bigg ) 
  e^{-(\lambda_0+\lambda_1+\lambda_2)D(\alpha,\tau^*)},
\end{eqnarray}
\noindent where $\delta_{i0}$, $\delta_{i1}$, $\delta_{i2}$, $n_0$, $n_1$, $n_2$ are same as defined before, 
$D(\alpha,\tau^*) = \sum_{i=1}^{n^*} t_{i:n}^{\alpha} + (n-n^*)\tau^*$.  Here also we assume same prior for $(\alpha, \lambda_0, \lambda_1, \lambda_2)$ for both the cases.  As before the posterior density can be written as below: \newline
In case of without order restricted inference
\begin{align}
 \begin{array}{lll} \label{posteri5}
  \widetilde{\pi}(\alpha, \lambda_0,\lambda_1,\lambda_2 \vert Data) & \propto & \widetilde{\pi_1}(\alpha) \widetilde{\pi_2}(\lambda_0,\lambda_1,\lambda_2 \vert \alpha),
 \end{array}
\end{align}
where, 
\begin{align}
 \begin{array}{lll}
  \widetilde{\pi_1}(\alpha) & = & e^{-c_1 \alpha} \alpha^{n^*+c_2-1} \big[b+D(\alpha,\tau^*) \big]^{-(a+n^*)} \prod_{i=1}^{n^*} t_{i:n}^{\alpha-1}, \nonumber \\
  \widetilde{\pi_2}(\lambda_0,\lambda_1,\lambda_2 \vert \alpha) & = & \frac{\Gamma(\overline{a}+n^*)}{\Gamma{(a+n^*)}}
  \big[\{b+D(\alpha,\tau^*)\} \lambda \big]^{[(a+n^*)-(\overline{a}+n^*)]} \\
  {} & {} & \times \prod_{j=0}^{2} \frac{\big[b+D(\alpha,\tau^*)\big]^{a_j+n_j}}{\Gamma(a_j+n_j)} 
  \lambda_j^{a_j+n_j-1} e^{-\lambda_j\big[b+D(\alpha,\tau^*)\big]}.
 \end{array}
\end{align}
In case of order restricted inference
\begin{align}
 \begin{array}{lll} \label{posteri6}
  \widetilde{\pi}(\alpha, \lambda_0,\lambda_1,\lambda_2 \vert Data) & \propto & \widetilde{\pi_1}(\alpha) \widetilde{\pi_2}(\lambda_0,\lambda_1,\lambda_2 \vert \alpha) h(\alpha, \lambda_0,\lambda_1,\lambda_2),
 \end{array}
\end{align}
where
\begin{align}
 \begin{array}{lll} \label{posteri7}
  \widetilde{\pi_1}(\alpha) & \propto & e^{-c_1 \alpha} \alpha^{n^*+c_2-1} \big[b+D(\alpha,\tau^*) \big]^{-(a+n^*)} \prod_{i=1}^{n^*} t_{i:n}^{\alpha-1}, \nonumber \\
  \widetilde{\pi_2}(\lambda_0,\lambda_1,\lambda_2 \vert \alpha) & = & \frac{\Gamma(\overline{a}+2n^*)}{\Gamma{(a+n^*)}\Gamma{(a_0+2n_0)}\Gamma{(a_1+n_1+n_2)}\Gamma{(a_2+n_1 + n_2)}}
  \big[\{b+D(\alpha,\tau^*) \} \lambda \big]^{[(a+n^*)-(\overline{a}+2n^*)]} \\
  {} & {} & \times \big[b+D(\alpha,\tau^*)\big]^{\overline{a}+2n^*} 
  \lambda_0^{a_0+2n_0-1} e^{-\lambda\big[b+D(\alpha,\tau^*)\big]} \\
	{} & {} & \times \big[ \lambda_1^{a_1+n_1 + n_2-1}\lambda_2^{a_2+n_1 + n_2-1} + \lambda_1^{a_2+n_1 + n_2-1}\lambda_2^{a_1+n_1 + n_2-1}\big], \\
h(\alpha, \lambda_0,\lambda_1,\lambda_2) & = & 	\frac{\lambda^{n^*}}{\lambda_0^{n_0} \lambda_1^{n_2} \lambda_2^{n_1}} .
 \end{array}
\end{align}
Now to obtain the Bayes estimates and the associated credible intervals, we can use Gibbs sampling technique in case of without order restricted inference and importance sampling technique in case of partially order restricted inference as explained in case of complete data.

\subsection{\sc Type-II Censoring}
In this censoring scheme the life testing experiment is terminated when the $r$-th (prefixed number) failure occurs, i.e, the total number of failure is fixed but the
termination time of the experiment is random. Available data under this censoring scheme is of the forms $\{(t_{1:n},\delta_1),\ldots,(t_{r:n},\delta_{r}) \}$.  Inference of Type-II censored data is very similar to that of Type-I censored data.  In this case we have to take $n^* = r$, $\tau^* = t_{r:n}$ and $D(\alpha,\tau^*) = \sum_{i=1}^{r} t_{i:n}^{\alpha} + (n-r)t_{r:n}$.  Also note that $r= n_0 + n_1 + n_2$.  All other expressions and the following analysis are same as the Type-I censoring scheme.

\subsection{\sc Type-I Hybrid Censoring}
The termination time in Type-I hybrid censoring scheme (HCS) is $\tau^* = min\{t_{r:n},\tau\}$, where $r$ is a pre-fixed number and $\tau$ is pre-fixed time.  If $n_1$ is the number of failures before $\tau$ then the available data under this censoring scheme is one of the following forms \newline
$(a)$ \quad $\{(t_{1:n},\delta_1),\ldots,(t_{n_1:n},\delta_{n_1}) \}$ if $\tau \leq t_{r:n},$ \newline
$(b)$ \quad $\{(t_{1:n},\delta_1),\ldots,(t_{r:n},\delta_{r}) \}$  if $t_{r:n}<\tau. $\newline 
Based on Type-I Hybrid censored data, the posterior analysis is same as that of Type-I censoring scheme with, for case (a) $n^* = n_1,$ $\tau^* = \tau,$ $D(\alpha,\tau^*) = \sum_{i=1}^{n_1} t_{i:n}^{\alpha} + (n-n_1)\tau$, and for case (b) $n^* = r$, $\tau^* = t_{r:n}$ and $D(\alpha,\tau^*) = \sum_{i=1}^{r} t_{i:n}^{\alpha} + (n-r)t_{r:n}$.  All other expressions and the following analysis are same as the Type-I censoring scheme.

\subsection{\sc Type-II Hybrid Censoring}
The termination time in Type-II HCS is $\tau^* = max\{t_{r:n},\tau\}$, where $r$ is a pre-fixed number and $\tau$ is pre-fixed time.  If $n_1$ is the number of failures before $\tau$ then the available data under this censoring scheme is one of the forms \newline
$(a)$ \quad $\{(t_{1:n},\delta_1),\ldots,(t_{r:n},\delta_{r}) \}$  if $\tau \leq t_{r:n},$ \newline
$(b)$ \quad $\{(t_{1:n},\delta_1),\ldots,(t_{n_1:n},\delta_{n_1}) \}$ if $t_{r:n}<\tau. $\newline 
Based on Type-II Hybrid censored data, the posterior analysis is same as that of Type-I censoring scheme with, for case (a) $n^* = r$, $\tau^* = t_{r:n}$ and $D(\alpha,\tau^*) = \sum_{i=1}^{r} t_{i:n}^{\alpha} + (n-r)t_{r:n}$, and for case (b) $n^* = n_1,$ $\tau^* = \tau,$ $D(\alpha,\tau^*) = \sum_{i=1}^{n_1} t_{i:n}^{\alpha} + (n-n_1)\tau$.  All other expressions and the following analysis are same as the Type-I censoring scheme.

\subsection{\sc Type-I Progressive Censoring}
Let $\tau_1, \ldots, \tau_k$ be $k$ pre-fixed time points and $R_1, \ldots, R_{k-1}$ be pre-fixed nonnegative integers less than $n$.  Also let $n_i$ $(i=1,\ldots, k)$ be the number of failures between time $\tau_{i-1}$ to $\tau_i$ $(\tau_0=0)$.  At the time $\tau_i$ $(i=1,\ldots, k-1)$, $R_i$ randomly chosen units from the survived units are removed from the experiment.  Finally $R_k = n - \sum_{i=1}^{k} n_i - \sum_{i=1}^{k-1} R_i$ units are removed at time $\tau_k$.  The available data in this censoring scheme is of the form $\{(t_{1:n},\delta_1),\ldots,(t_{n^*:n},\delta_{n^*}) \}$.  Based on Type-I progressive censored data, the posterior analysis is same as that of Type-I censoring scheme with, $n^* = \sum_{i=1}^{k} n_i$, $\tau^* = \tau_k$ and $D(\alpha,\tau^*) = \sum_{i=1}^{n^*} t_{i:n}^{\alpha} + \sum_{i=1}^{k} R_i \tau_i^{\alpha}$.  All other expressions and the following analysis are same as the Type-I censoring scheme. 

\subsection{\sc Type-II Progressive Censoring}
Let $R_1, \ldots, R_{m}$ be pre-fixed nonnegative integers such that $m+\sum_{i=1}^{m} R_i = n$.  Under this censoring scheme, at the time of first failure, say $t_{i:n}$, $R_1$ randomly chosen experimental units from the remaining $n-1$ are removed from the experiment. Similarly at the time of second failure, say $t_{2:n}$, $R_2$ randomly chosen experimental units from the remaining $n-R_1-2$ units are removed from the experiment and finally at the time of $m$-th failure, say $t_{m:n}$, all the remaining $R_m$ units are removed from the experiment.  The available data in this censoring scheme is of the form $\{(t_{1:n},\delta_1),\ldots,(t_{m:n},\delta_{m}) \}$.  Based on Type-II progressive censored data, the posterior analysis is same as that of Type-I censoring scheme with, $n^* = m$, $\tau^* = t_{m:n}$ and $D(\alpha,\tau^*) = \sum_{i=1}^{m} (R_i + 1) t_{i:n}^{\alpha} $.  All other expressions and the following analysis are same as the Type-I censoring scheme.    

\section{\sc Testing of Hypothesis}
\label{sec:test}
In this section we provide a method of testing the hypothesis that both the causes have equal effect.  Mathematically, we want to test the null hypothesis $H_0 : \lambda_1 = \lambda_2$ against the alternative $H_1 : \lambda_1 \neq \lambda_2$.  Therefore under $H_0$, i.e. under the assumption of equality of two causes of failure we may assume that the data ${\bf t} = (t_{1:n}, \ldots, t_{n:n})$ is coming from a Weibull distribution with parameters $\alpha^*$ and $\lambda^*$.  We propose to use Bayes factor for testing the hypothesis.  Under $H_1$, the likelihood function and the joint prior distribution are given in equation (\ref{likeli2}) and equation (\ref{prior2}) respectively.  Under $H_0$, the likelihood function is given by 
\begin{align}
\begin{array}{lll}
L_1({\bf t} \vert \alpha^*, \lambda^*) & = & \alpha^{*n} \lambda^{*n} e^{-\lambda^* \sum_{i=1}^{n} t_{i:n}^{\alpha^*}} \prod_{i=1}^{n}  t_{i:n}^{\alpha^*-1} .
\end{array}
\end{align}  
Assume that the prior distributions of $\alpha^{*}$ and $\lambda^{*}$ are $GA(d_1,d_2)$ and $GA(d_3,d_4)$ respectively.  Also assume that the prior distributions of $\alpha^{*}$ and $\lambda^{*}$ are independent.  Hence the joint density function of $\alpha^{*}$ and $\lambda^{*}$ is
\begin{align}
\begin{array}{lll}
\pi(\alpha^*, \lambda^*) & = & \frac{d_1^{d_2}}{\Gamma(d_2)} e^{-d_1 \alpha^*} \alpha^{* d_2-1} \times 
\frac{d_3^{d_4}}{\Gamma(d_4)} e^{-d_3 \lambda^*} \lambda^{* d_4-1}.
\end{array}
\end{align}
Therefore, under $H_0$, the marginal distribution of ${\bf t}$ is given by
 \begin{align}
\begin{array}{lll}
l_{H_0}({\bf t}) & = & \displaystyle \int_{0}^{\infty} \int_{0}^{\infty} L_1({\bf t} \vert \alpha^*, \lambda^*)  \pi(\alpha^*, \lambda^*) d \alpha^* d \lambda^* \\
\\
{} & = & \displaystyle \frac{\Gamma (n+d_4) d_1^{d_2} d_3^{d_4}}{A \Gamma(d_2) \Gamma(d_4) } ,
\end{array}
\end{align}
where
 \begin{align}
\begin{array}{lll}
\frac{1}{A} & = & \displaystyle \int_{0}^{\infty} \alpha^{* n + d_2-1} e^{-d_1 \alpha^*}  (d_3 + \sum_{i=1}^{n} t_{i:n}^{\alpha^*})^{-(n+d_4)} \prod_{i=1}^{n}  t_{i:n}^{\alpha^*-1} d \alpha^*. 
\end{array}
\end{align}
Similarly the marginal distribution of ${\bf t}$ under $H_1$ is given by
 \begin{align}
\begin{array}{lll}
l_{H_1}({\bf t}) & = & \displaystyle \int_{0}^{\infty} \ldots \int_{0}^{\infty} L(\alpha, \lambda_0, \lambda_1, \lambda_2 \vert Data) \pi_1 (\alpha,\lambda_0, \lambda_1, \lambda_2 \vert a, b, a_0, a_1, a_2,c_1,c_2) d \alpha d \lambda_0 d \lambda_1 d \lambda_2 \\
\\
{} & = & \displaystyle \frac{\Gamma (n_0+a_0) \Gamma (n_1+a_1) \Gamma (n_2+a_2) \Gamma (n+a) \Gamma (\overline{a}) b^{a} c_1^{c_2}}{A \Gamma (a_0) \Gamma (a_1) \Gamma (a_2) \Gamma (n+\overline{a}) \Gamma(a) \Gamma(c_2) } .
\end{array}
\end{align}
Therefore the Bayes factor (BF) for testing $H_0$ against $H_1$ is
$$
BF  =  \frac{l_{H_0}({\bf t})}{l_{H_1}({\bf t})} 
 =  \frac{d_1^{d_2} d_3^{d_4} \Gamma (n+d_4)  \Gamma (a_0) \Gamma (a_1) \Gamma (a_2) \Gamma (n+\overline{a}) \Gamma(a) \Gamma(c_2)}{ b^{a} c_1^{c_2} \Gamma (n_0+a_0) \Gamma (n_1+a_1) \Gamma (n_2+a_2) \Gamma (n+a) \Gamma (\overline{a})  \Gamma(d_2) \Gamma(d_4) } .
$$
Hence for given data, we reject $H_0$ if BF is low.  We illustrate this testing of hypothesis in data analysis section.

\section{\sc Data Analysis}
Diabetic Retinopathy is one of the major causes of vision loss and blindness of diabetes patients.  National Eye
Institute conducted DRS to estimate the effect of laser treatment in reducing the risk of blindness.  The study
was conducted on $71$ patients.  For each patient, one eye was selected at random and the laser treatment was given on that eye.  For each patient the time to blindness and the indicator mentioning whether treated or untreated or
both eyes became blind has been recorded.  The main purpose of this study is to verify whether the laser treatment
has any effect in delaying the onset of blindness in patients with diabetic retinopathy.
The treatment or lack of treatment can be regarded as two causes of blindness, hence this data set can be treated
as a competing risks data.  Clearly, the two competing causes in this case cannot be taken as independent.
Moreover, there is a positive probability of simultaneous occurrence of both the causes.
Hence, MOBW distribution is a plausible model to analyze this data set.  

We have analyzed the data after dividing the failure time by 365, i.e., by changing the unit of failure time from
day to year.  It is not going to affect the conclusions of the study.
We have provided the Bayesian inference of the model parameters.  Since we do not have any prior information
on the model parameters, we have assumed proper priors which are almost non-informative as suggested by
Congdon (2003).  The hyper parameters are $a=b=c_1=c_2=0.001$ and $a_0=a_1=a_2=1$.

The Bayes estimates of $\alpha$, $\lambda_0$, $\lambda_1$ and $\lambda_2$ without assuming any order restriction
are $1.5393$, $0.0714$, $0.1872$ and $0.2207$ respectively.  The symmetric and HPD credible intervals without assuming order restriction are provided in Table \ref{tab:data1CRI1}.  Next we analyze the data assuming $\lambda_1 < \lambda_2$, i.e., the expected time to blindness of the treated eye is higher than the eye without the laser treatment.  The Bayes estimates of $\alpha$, $\lambda_0$, $\lambda_1$ and $\lambda_2$ assuming order restriction are $1.5388$, $0.0707$, $0.1789$ and $0.2281$, respectively.  The symmetric and HPD credible intervals assuming order restriction are provided in Table \ref{tab:data1CRI2}.

\begin{table}[!ht]\scriptsize
\caption{Diabetic Retinopathy Data.}
\centering
\begin{tabular}{crr*{10}{r}}
\toprule
 $i$        & 1   & 2    & 3    & 4    & 5   & 6    & 7   & 8   & 9   & 10   & 11  & 12 \\
 $t_{i:n}$  & 266 & 91   & 154  & 285  & 583 & 547  & 79  & 622 & 707 & 469  & 93  & 1313 \\
 $\delta_i$ & 1   & 2    & 2    & 0    & 1   & 2    & 1   & 0   & 2   & 2    & 1    & 2  \\
\\
\midrule
 $i$        & 13  & 14   & 15   & 16   & 17  & 18   & 19  & 20  & 21  & 22   & 23  & 24 \\
 $t_{i:n}$  & 805 & 344  & 790  & 125  & 777 & 306  & 415 & 307 & 637 & 577  & 178 & 517 \\
 $\delta_i$ & 1   & 1    & 2    & 2    & 2   & 1    & 1   & 2   & 2   & 2    & 1   & 2   \\
\\
\midrule
 $i$        & 25  & 26   & 27   & 28   & 29  & 30   & 31  & 32  & 33  & 34   & 35  & 36 \\
 $t_{i:n}$  & 272 & 1137 & 1484 & 315  & 287 & 1252 & 717 & 642 & 141 & 407  & 356 & 1653 \\
 $\delta_i$ & 0   & 0    & 1    & 1    & 2   & 1    & 2   & 1   & 2   & 1    & 1   & 0    \\
\\
\midrule
 $i$        & 37  & 38   & 39   & 40   & 41  & 42   & 43  & 44  & 45  & 46   & 47  & 48 \\
 $t_{i:n}$  & 427 & 699  & 36   & 667  & 588 & 471  & 126 & 350 & 350 & 663  & 567 & 966  \\
 $\delta_i$ & 2   & 1    & 2    & 1    & 2   & 0    & 1   & 2   & 1   & 0    & 2   & 0    \\
\\
\midrule
 $i$        & 49  & 50   & 51   & 52   & 53  & 54   & 55  & 56  & 57  & 58   & 59  & 60   \\
 $t_{i:n}$  & 203 & 84   & 392  & 1140 & 901 & 1247 & 448 & 904 & 276 & 520  & 485 & 248 \\
 $\delta_i$ & 0   & 1    & 1    & 2    & 1   & 0    & 2   & 2   & 1   & 1    & 2   & 2   \\
\\
\midrule
 $i$        & 61  & 62   & 63   & 64   & 65  & 66   & 67  & 68  & 69  & 70   & 71  & {}  \\
 $t_{i:n}$  & 503 & 423  & 285  & 315  & 727 & 210  & 409 & 584 & 355 & 1302 & 227 & {} \\
 $\delta_i$ & 1   & 2    & 2    & 2    & 2   & 2    & 2   & 1   & 1   & 1    & 2   & {} \\
\bottomrule
\end{tabular}
\label{tab:data1}
\end{table}

\begin{table}[!ht]\scriptsize
\caption{Symmetric and HPD CRIs of diabetic retinopathy data set (Without order restriction).}
\centering
\begin{tabular}{ll*{10}{r}}
\toprule
 {} & {} & \multicolumn{2}{c}{$\alpha$} & \multicolumn{2}{c}{$\lambda_0$} & \multicolumn{2}{c}{$\lambda_1$} & \multicolumn{2}{c}{$\lambda_2$}   \\  
\cmidrule(lr){3-4}\cmidrule(lr){5-6}\cmidrule(lr){7-8}\cmidrule(lr){9-10}
CI & Level & \multicolumn{1}{c}{LL} & \multicolumn{1}{c}{UL} & \multicolumn{1}{c}{LL} & \multicolumn{1}{c}{UL} & \multicolumn{1}{c}{LL} & \multicolumn{1}{c}{UL} & \multicolumn{1}{c}{LL} & \multicolumn{1}{c}{UL}  \\
\midrule
{}          & 90\% & 1.2562 & 1.8847 & 0.0369 & 0.1159 & 0.1199 & 0.2675 & 0.1435 & 0.3123  \\
Symmetric   & 95\% & 1.2244 & 1.9261 & 0.0324 & 0.1270 & 0.1113 & 0.2853 & 0.1336 & 0.3313  \\
{}          & 99\% & 1.1611 & 2.0080 & 0.0249 & 0.1527 & 0.0932 & 0.3211 & 0.1160 & 0.3694  \\
\midrule
{}          & 90\% &  1.2518 & 1.8773 & 0.0310 & 0.1079 & 0.1127 & 0.2567 & 0.1334 & 0.2989  \\
HPD         & 95\% &  1.2167 & 1.9139 & 0.0271 & 0.1186 & 0.1054 & 0.2772 & 0.1276 & 0.3210  \\
{}          & 99\% &  1.1568 & 1.9979 & 0.0203 & 0.1434 & 0.0892 & 0.3123 & 0.1108 & 0.3593  \\
\bottomrule
\end{tabular}
\label{tab:data1CRI1}
\end{table}

\begin{table}[!ht]\scriptsize
\caption{Symmetric and HPD CRIs of diabetic retinopathy data set (Order restricted). }
\centering
\begin{tabular}{ll*{10}{r}}
\toprule
 {} & {} & \multicolumn{2}{c}{$\alpha$} & \multicolumn{2}{c}{$\lambda_0$} & \multicolumn{2}{c}{$\lambda_1$} & \multicolumn{2}{c}{$\lambda_2$}   \\  
\cmidrule(lr){3-4}\cmidrule(lr){5-6}\cmidrule(lr){7-8}\cmidrule(lr){9-10}
CI & Level & \multicolumn{1}{c}{LL} & \multicolumn{1}{c}{UL} & \multicolumn{1}{c}{LL} & \multicolumn{1}{c}{UL} & \multicolumn{1}{c}{LL} & \multicolumn{1}{c}{UL} & \multicolumn{1}{c}{LL} & \multicolumn{1}{c}{UL}  \\
\midrule
{}          & 90\% &  1.2525 & 1.8805 & 0.0371 & 0.1134 & 0.1201 & 0.2458 & 0.1580 & 0.3097  \\
Symmetric   & 95\% &  1.2189 & 1.9137 & 0.0330 & 0.1243 & 0.1114 & 0.2612 & 0.1478 & 0.3278  \\
{}          & 99\% &  1.1508 & 1.9785 & 0.0261 & 0.1448 & 0.0977 & 0.2881 & 0.1296 & 0.3598  \\
\midrule
{}          & 90\% &  1.2512 & 1.8769 & 0.0306 & 0.1039 & 0.1155 & 0.2394 & 0.1535 & 0.3026  \\
HPD         & 95\% &  1.2123 & 1.9043 & 0.0306 & 0.1165 & 0.1065 & 0.2527 & 0.1432 & 0.3207  \\
{}          & 99\% &  1.1475 & 1.9746 & 0.0228 & 0.1333 & 0.0938 & 0.2782 & 0.1258 & 0.3543  \\
\bottomrule
\end{tabular}
\label{tab:data1CRI2}
\end{table}

\begin{figure}[h]
\begin{center}
\includegraphics[width=4in,height=2.5in,angle=0]{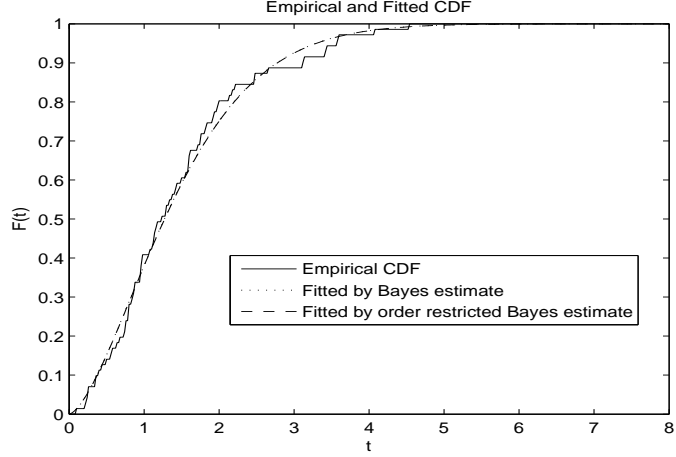}
\caption{Plot of Empirical CDF and Fitted CDF for diabetic retinopathy data set.}
\label{fig:empfitted}
\end{center}
\end{figure}

Next we have checked the goodness of fit of the data using Kolmogorov-Smirnov (KS) test statistics.  
The KS distance between empirical and fitted CDF using Bayes estimates without order restriction is 
$0.0579$, which indicates that the empirical and fitted CDF are very close.  The $p$- value of the test for testing the equality of two CDFs is $0.9598$, i.e., based on the data we cannot reject the hypothesis that the empirical and fitted CDF are equal.  We have performed the test for order restricted case also.  The KS distance and $p$- value in this case are $0.0572$ and $0.9637$ respectively.  Note that the KS distance in case of order restricted inference is smaller than the without order restricted inference.  The graphical representation of empirical and fitted CDF is provided in Figure \ref{fig:empfitted}.

Next we have tested the hypothesis that there is no significance difference between two causes of failure, i.e., we have tested the null hypothesis $H_0 : \lambda_1 = \lambda_2$ against the alternative $H_1 : \lambda_1 \neq \lambda_2$ using the method proposed in Section \ref{sec:test}.  The Bayes factor for the given data using almost non-informative prior is $2.326479\times 10^{32}$.  Since the BF is very high, we conclude that two causes are not significantly
different.  Therefore, the conclusion from the study is that the laser treatment does not have any effect in
delaying the onset of blindness to the patients with diabetic retinopathy.

\begin{figure}[h]
\begin{center}
\includegraphics[width=4in,height=2.5in,angle=0]{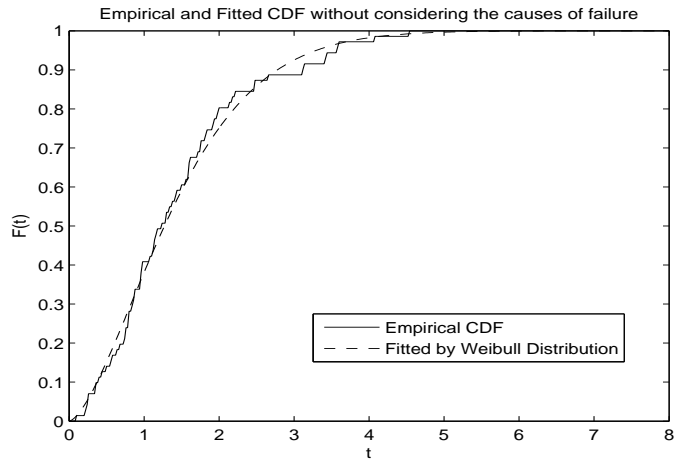}
\caption{Plot of Empirical CDF and Fitted CDF assuming equality of two causes of failure.}
\label{fig:empfitted2}
\end{center}
\end{figure}

Now, we consider the data without the causes of failure and fitted the data assuming Weibull distribution with shape parameter $\alpha^*$ and scale parameter $\lambda^*$.  Assuming almost non-informative gamma prior for both $\alpha^*$ and $\lambda^*$, the Bayes estimates of 
$\alpha^*$ and $\lambda^*$ are respectively $1.5358$ and $0.4795$.  The KS distance and p-value of the fit are respectively $0.0582$ and $0.9581$, which indicates the good fit of the data.  The empirical and the fitted CDF of the data assuming Weibull distribution is presented in Figure \ref{fig:empfitted2}.   This can be used to estimate
$E(T)$, i.e. the expected time to the onset of blindness, or to estimate $E(T|T > a)$, for some $a > 0$, etc.

\section{\sc Simulation}
In this section we provide an extensive simulation study based on complete sample to verify how the proposed
estimators behave for different sample sizes and for different set of parameters.  Simulation results are
provided for both, without order restriction and with order restriction on scale parameters.  We consider three sets of parameter values of $(\alpha, \, \lambda_0, \, \lambda_1, \, \lambda_2)$: Set I $(2.0, 0.5, 1.0, 1.2)$, Set II $(2.0, 1.0, 1.0, 1.2)$ and Set III $(2.0, 1.5, 1.0, 1.2)$.  We have taken $n$ = 30, 40, 50.  We have considered almost non-informative proper priors, as suggested by 
 Congdon (2003); the hyper parameters are $a=b=c_1=c_2=0.001$
and $a_0=a_1=a_2=1$.  We provide the average estimates (AEs) along with the mean square errors (MSEs) of the model parameters.  The average lengths (AL) and coverage percentages (CP) of $95\%$ symmetric and highest posterior density (HPD) credible intervals are also provided.  All the simulation results are provided from Table \ref{tab:AE1} to Table \ref{tab:CRI4} and the results are based on 5000 replications.

Some of the points are very clear from the simulation experiments.  Both biases and MSEs are decreases with the increase of sample size $n$ and hence it indicates the consistency of the estimator.  If we observe the ALs and CPs of different credible intervals, in all the cases CPs are closed to the nominal values and ALs are decreases with the increase of $n$.  Now if we compare the inference based on partially order restriction on scale parameters with unrestricted inference then it has been observed that order restricted inference provides lower MSEs for $\lambda_1$ and $\lambda_2$ than unrestricted inference.  Also the ALs of different CRIs of $\lambda_1$ and $\lambda_2$ under order restricted inference is is lower than that of unrestricted inference.

\begin{table}[!ht]\scriptsize
\caption{Without order restricted Bayes estimates along with the corresponding mean square errors ($\alpha=2.0$, $\lambda_1=1.0$, $\lambda_2 = 1.2$).}
\centering
\begin{tabular}{*{2}{c}*{8}{r}}
\toprule
\multicolumn{2}{c}{} & \multicolumn{2}{c}{$\alpha$} & \multicolumn{2}{c}{$\lambda_0$} & \multicolumn{2}{c}{$\lambda_1$} & \multicolumn{2}{c}{$\lambda_2$}\\
\cmidrule(lr){3-4} \cmidrule(lr){5-6} \cmidrule(lr){7-8} \cmidrule(lr){9-10}
n & $\lambda_0$  & \multicolumn{1}{c}{AE} & \multicolumn{1}{c}{MSE} & \multicolumn{1}{c}{AE} & \multicolumn{1}{c}{MSE} & \multicolumn{1}{c}{AE} & \multicolumn{1}{c}{MSE}  & \multicolumn{1}{c}{AE} & \multicolumn{1}{c}{MSE} \\
\midrule
30 & 0.5 & 2.035 & 0.091 & 0.569 & 0.062 & 1.050 & 0.120 & 1.232 & 0.150  \\
40 & 0.5 & 2.027 & 0.067 & 0.546 & 0.039 & 1.033 & 0.083 & 1.226 & 0.101  \\
50 & 0.5 & 2.013 & 0.053 & 0.538 & 0.032 & 1.034 & 0.068 & 1.214 & 0.079  \\
\\
30 & 1.0 & 2.031 & 0.091 & 1.071 & 0.156 & 1.074 & 0.162 & 1.270 & 0.194  \\
40 & 1.0 & 2.023 & 0.067 & 1.054 & 0.108 & 1.058 & 0.105 & 1.241 & 0.131  \\
50 & 1.0 & 2.018 & 0.052 & 1.047 & 0.085 & 1.042 & 0.081 & 1.240 & 0.101  \\
\\
30 & 1.5 & 2.036 & 0.091 & 1.608 & 0.375 & 1.108 & 0.215 & 1.316 & 0.256  \\
40 & 1.5 & 2.019 & 0.066 & 1.562 & 0.212 & 1.077 & 0.139 & 1.271 & 0.167  \\
50 & 1.5 & 2.015 & 0.050 & 1.546 & 0.156 & 1.056 & 0.094 & 1.254 & 0.118  \\
\bottomrule 
\end{tabular}\label{tab:AE1}
\end{table}

\begin{table}[!ht]\scriptsize
\caption{Order restricted Bayes estimates along with the corresponding mean square errors ($\alpha=2.0$, $\lambda_1=1.0$, $\lambda_2 = 1.2$).}
\centering
\begin{tabular}{*{2}{c}*{8}{r}}
\toprule
\multicolumn{2}{c}{} & \multicolumn{2}{c}{$\alpha$} & \multicolumn{2}{c}{$\lambda_0$} & \multicolumn{2}{c}{$\lambda_1$} & \multicolumn{2}{c}{$\lambda_2$}\\
\cmidrule(lr){3-4} \cmidrule(lr){5-6} \cmidrule(lr){7-8} \cmidrule(lr){9-10}
n & $\lambda_0$  & \multicolumn{1}{c}{AE} & \multicolumn{1}{c}{MSE} & \multicolumn{1}{c}{AE} & \multicolumn{1}{c}{MSE} & \multicolumn{1}{c}{AE} & \multicolumn{1}{c}{MSE}  & \multicolumn{1}{c}{AE} & \multicolumn{1}{c}{MSE} \\
\midrule
30 & 0.5 & 2.041 & 0.095 & 0.570 & 0.062 & 0.939 & 0.066 & 1.352 & 0.147  \\
40 & 0.5 & 2.033 & 0.069 & 0.545 & 0.041 & 0.949 & 0.046 & 1.317 & 0.097  \\
50 & 0.5 & 2.016 & 0.052 & 0.537 & 0.033 & 0.952 & 0.036 & 1.286 & 0.065  \\
\\
30 & 1.0 & 2.041 & 0.096 & 1.081 & 0.166 & 0.950 & 0.084 & 1.406 & 0.204  \\ 
40 & 1.0 & 2.024 & 0.068 & 1.051 & 0.107 & 0.959 & 0.063 & 1.360 & 0.135  \\
50 & 1.0 & 2.022 & 0.055 & 1.039 & 0.081 & 0.959 & 0.045 & 1.321 & 0.091  \\
\\
30 & 1.5 & 2.040 & 0.096 & 1.615 & 0.350 & 0.964 & 0.114 & 1.453 & 0.289  \\
40 & 1.5 & 2.027 & 0.069 & 1.567 & 0.216 & 0.957 & 0.071 & 1.383 & 0.163  \\
50 & 1.5 & 2.015 & 0.052 & 1.552 & 0.168 & 0.960 & 0.055 & 1.346 & 0.115  \\
\bottomrule 
\end{tabular}\label{tab:AE2}
\end{table}

\begin{table}[!ht]\scriptsize
\caption{Coverage percentage and average length of $95\%$ symmetric CRIs (Without order restricted, $\alpha=2.0$, $\lambda_1=1.0$, $\lambda_2 = 1.2$).}
\centering
\begin{tabular}{*{2}{c}*{8}{r}}
\toprule
\multicolumn{2}{c}{} & \multicolumn{2}{c}{$\alpha$} & \multicolumn{2}{c}{$\lambda_0$} & \multicolumn{2}{c}{$\lambda_1$} & \multicolumn{2}{c}{$\lambda_2$}\\
\cmidrule(lr){3-4} \cmidrule(lr){5-6} \cmidrule(lr){7-8} \cmidrule(lr){9-10}
n & $\lambda_0$  & \multicolumn{1}{c}{AL} & \multicolumn{1}{c}{CP} & \multicolumn{1}{c}{ALL} & \multicolumn{1}{c}{CP} & \multicolumn{1}{c}{AL} & \multicolumn{1}{c}{CP}  & \multicolumn{1}{c}{AL} & \multicolumn{1}{c}{CP} \\
\midrule
30 & 0.5 &  1.441 & 98.54 & 0.920 & 95.72 & 1.331 & 96.08 & 1.472 & 95.78  \\
40 & 0.5 &  1.257 & 98.68 & 0.777 & 96.30 & 1.133 & 95.70 & 1.259 & 95.70  \\
50 & 0.5 &  1.124 & 98.62 & 0.691 & 95.18 & 1.013 & 95.38 & 1.116 & 95.84  \\
\\
30 & 1.0 &  1.437 & 98.38 & 1.518 & 95.96 & 1.522 & 96.10 & 1.695 & 96.18  \\
40 & 1.0 &  1.253 & 98.30 & 1.290 & 95.92 & 1.294 & 96.64 & 1.433 & 95.98  \\
50 & 1.0 &  1.128 & 98.74 & 1.149 & 96.00 & 1.145 & 96.16 & 1.278 & 95.98  \\
\\
30 & 1.5 &  1.442 & 98.34 & 2.231 & 96.08 & 1.733 & 96.02 & 1.941 & 96.50  \\
40 & 1.5 &  1.250 & 98.70 & 1.850 & 96.12 & 1.447 & 96.36 & 1.612 & 96.28  \\
50 & 1.5 &  1.126 & 98.76 & 1.631 & 96.84 & 1.270 & 96.62 & 1.420 & 96.58  \\
\bottomrule 
\end{tabular}\label{tab:CRI1}
\end{table}

\begin{table}[!ht]\scriptsize
\caption{Coverage percentage and average length of $95\%$ HPD CRIs (Without order restricted, $\alpha=2.0$, $\lambda_1=1.0$, $\lambda_2 = 1.2$).}
\centering
\begin{tabular}{*{2}{c}*{8}{r}}
\toprule
\multicolumn{2}{c}{} & \multicolumn{2}{c}{$\alpha$} & \multicolumn{2}{c}{$\lambda_0$} & \multicolumn{2}{c}{$\lambda_1$} & \multicolumn{2}{c}{$\lambda_2$}\\
\cmidrule(lr){3-4} \cmidrule(lr){5-6} \cmidrule(lr){7-8} \cmidrule(lr){9-10}
n & $\lambda_0$  & \multicolumn{1}{c}{AL} & \multicolumn{1}{c}{CP} & \multicolumn{1}{c}{ALL} & \multicolumn{1}{c}{CP} & \multicolumn{1}{c}{AL} & \multicolumn{1}{c}{CP}  & \multicolumn{1}{c}{AL} & \multicolumn{1}{c}{CP} \\
\midrule
30 & 0.5  & 1.434 & 98.48 & 0.871 & 94.96 & 1.284 & 95.12 & 1.423 & 94.68  \\
40 & 0.5  & 1.251 & 98.64 & 0.745 & 95.50 & 1.103 & 94.86 & 1.228 & 95.00  \\
50 & 0.5  & 1.121 & 98.58 & 0.668 & 94.88 & 0.991 & 94.96 & 1.093 & 94.84  \\
\\
30 & 1.0  & 1.430 & 98.46 & 1.449 & 95.02 & 1.453 & 95.14 & 1.624 & 95.38  \\
40 & 1.0  & 1.248 & 98.26 & 1.246 & 95.60 & 1.249 & 96.02 & 1.387 & 95.22  \\
50 & 1.0  & 1.124 & 98.76 & 1.116 & 95.72 & 1.113 & 95.58 & 1.245 & 95.72  \\
\\
30 & 1.5  & 1.435 & 98.28 & 2.124 & 95.80 & 1.636 & 95.40 & 1.841 & 95.48  \\
40 & 1.5  & 1.245 & 98.72 & 1.785 & 95.62 & 1.387 & 95.48 & 1.549 & 95.68  \\
50 & 1.5  & 1.122 & 98.74 & 1.584 & 95.82 & 1.227 & 96.10 & 1.375 & 96.26  \\
\bottomrule 
\end{tabular}\label{tab:CRI2}
\end{table}

\begin{table}[!ht]\scriptsize
\caption{Coverage percentage and average length of $95\%$ symmetric CRIs (Order restricted, $\alpha=2.0$, $\lambda_1=1.0$, $\lambda_2 = 1.2$).}
\centering
\begin{tabular}{*{2}{c}*{8}{r}}
\toprule
\multicolumn{2}{c}{} & \multicolumn{2}{c}{$\alpha$} & \multicolumn{2}{c}{$\lambda_0$} & \multicolumn{2}{c}{$\lambda_1$} & \multicolumn{2}{c}{$\lambda_2$}\\
\cmidrule(lr){3-4} \cmidrule(lr){5-6} \cmidrule(lr){7-8} \cmidrule(lr){9-10}
n & $\lambda_0$  & \multicolumn{1}{c}{AL} & \multicolumn{1}{c}{CP} & \multicolumn{1}{c}{ALL} & \multicolumn{1}{c}{CP} & \multicolumn{1}{c}{AL} & \multicolumn{1}{c}{CP}  & \multicolumn{1}{c}{AL} & \multicolumn{1}{c}{CP} \\
\midrule
30 & 0.5  & 1.445 & 98.14 & 0.886 & 94.24 & 1.024 & 93.94 & 1.388 & 97.22  \\
40 & 0.5  & 1.261 & 98.12 & 0.744 & 94.56 & 0.885 & 95.12 & 1.162 & 97.22  \\
50 & 0.5  & 1.126 & 98.76 & 0.658 & 93.88 & 0.787 & 95.26 & 1.011 & 97.34  \\
\\
30 & 1.0  & 1.447 & 98.44 & 1.496 & 95.16 & 1.174 & 94.50 & 1.628 & 97.22  \\
40 & 1.0  & 1.253 & 98.04 & 1.255 & 95.56 & 1.013 & 95.44 & 1.354 & 96.72  \\
50 & 1.0  & 1.130 & 98.44 & 1.105 & 95.66 & 0.899 & 95.74 & 1.172 & 97.42  \\
\\
30 & 1.5  & 1.445 & 98.32 & 2.190 & 95.94 & 1.323 & 95.04 & 1.863 & 97.08  \\ 
40 & 1.5  & 1.256 & 98.26 & 1.814 & 95.76 & 1.118 & 95.56 & 1.516 & 97.32  \\
50 & 1.5  & 1.125 & 98.62 & 1.598 & 95.60 & 0.996 & 96.04 & 1.318 & 97.72  \\
\bottomrule 
\end{tabular}\label{tab:CRI3}
\end{table}

\begin{table}[!ht]\scriptsize
\caption{Coverage percentage and average length of $95\%$ HPD CRIs (Partially Order restricted, $\alpha=2.0$, $\lambda_1=1.0$, $\lambda_2 = 1.2$).}
\centering
\begin{tabular}{*{2}{c}*{8}{r}}
\toprule
\multicolumn{2}{c}{} & \multicolumn{2}{c}{$\alpha$} & \multicolumn{2}{c}{$\lambda_0$} & \multicolumn{2}{c}{$\lambda_1$} & \multicolumn{2}{c}{$\lambda_2$}\\
\cmidrule(lr){3-4} \cmidrule(lr){5-6} \cmidrule(lr){7-8} \cmidrule(lr){9-10}
n & $\lambda_0$  & \multicolumn{1}{c}{AL} & \multicolumn{1}{c}{CP} & \multicolumn{1}{c}{ALL} & \multicolumn{1}{c}{CP} & \multicolumn{1}{c}{AL} & \multicolumn{1}{c}{CP}  & \multicolumn{1}{c}{AL} & \multicolumn{1}{c}{CP} \\
\midrule

30 & 0.5  & 1.433 & 98.02 & 0.834 & 93.56 & 0.984 & 91.54 & 1.338 & 97.70  \\
40 & 0.5  & 1.251 & 98.20 & 0.709 & 93.98 & 0.856 & 92.80 & 1.128 & 97.60  \\
50 & 0.5  & 1.118 & 98.78 & 0.631 & 93.20 & 0.766 & 93.04 & 0.984 & 97.58  \\
\\
30 & 1.0  & 1.434 & 98.34 & 1.415 & 94.24 & 1.117 & 91.58 & 1.557 & 98.18  \\
40 & 1.0  & 1.243 & 97.94 & 1.201 & 94.56 & 0.974 & 93.34 & 1.305 & 98.16  \\
50 & 1.0  & 1.122 & 98.30 & 1.065 & 94.80 & 0.869 & 93.48 & 1.136 & 98.08  \\
\\
30 & 1.5  & 1.433 & 98.36 & 2.069 & 94.72 & 1.246 & 92.22 & 1.767 & 98.50  \\
40 & 1.5  & 1.246 & 98.30 & 1.735 & 94.90 & 1.066 & 93.02 & 1.455 & 98.24  \\
50 & 1.5  & 1.117 & 98.62 & 1.540 & 95.04 & 0.957 & 93.72 & 1.272 & 98.16  \\
\bottomrule 
\end{tabular}\label{tab:CRI4}
\end{table}

\section{\sc Conclusion}
In this article we have provided the Bayesian inference of a dependent competing risk model.  We assume Marshall-Olkin bivariate Weibull distribution to explain the dependency structure between two causes of failure.  Bayesian inference has been provided under two scenario, in one case we assume order restriction between two causes of failures and in other case we do not assume any order restriction.  We have also shown that the inference procedure for both the cases can easily be extended to different censoring schemes.  We propose to use Bayes factor to test the hypothesis that there is no significant difference between two causes of failure.  An extensive simulation results and a data analysis shows that the proposed method works quite well.  If it is known apriori that one cause of failure is higher risk than the other then it is better to use the order restricted inference.




\section*{\sc Appendix}
\noindent {\sc Proof of Theorem 1:}

\begin{align}
\begin{array}{lll}
\ln(\widetilde{\pi_1}(\alpha)) &= & -c_1 \alpha + (n+c_2-1) \ln (\alpha) - (a+n) \ln (b + \sum_{i=1}^{n} t_{i:n}^{\alpha}) + (\alpha - 1) \sum_{i=1}^{n} \ln (t_{i:n}), \nonumber \\
\frac{\partial^2 \ln(\widetilde{\pi_1}(\alpha))}{\partial \alpha^2} &= & - \frac{n+c_2-1}{\alpha^2} 
- (a+n)\bigg[\frac{b \sum_{i=1}^{n} t_{i:n}^{\alpha}(\ln (t_{i:n}))^2 + \sum_{i=1}^{n} t_{i:n}^{\alpha} \sum_{i=1}^{n} t_{i:n}^{\alpha}(\ln (t_{i:n}))^2 - (\sum_{i=1}^{n} t_{i:n}^{\alpha}\ln (t_{i:n}))^2 }{(b + \sum_{i=1}^{n} t_{i:n}^{\alpha})^2 }\bigg]\\ {} & \leq & 0. \nonumber
\end{array}
\end{align} 
Since, $\sum_{i=1}^{n} t_{i:n}^{\alpha} \sum_{i=1}^{n} t_{i:n}^{\alpha}(\ln (t_{i:n}))^2 - (\sum_{i=1}^{n} t_{i:n}^{\alpha}\ln (t_{i:n}))^2 \geq 0$ (by Cauchy-Schwarz inequality). 

\end{document}